\documentclass[apjl]{emulateapj}
\usepackage{mathptmx}

\begin{document}

\title{Cosmic Analogues of the Stern-Gerlach Experiment and the Detection of Light Bosons}
\author{Doron Chelouche\altaffilmark{1} and Eduardo I. Guendelman\altaffilmark{2}}
\altaffiltext{1} {Canadian Institute for Theoretical Astrophysics, 60 St. George st., Toronto ON M4W 5H8, Canada; doron@cita.utoronto.ca}
\altaffiltext{2} {Physics Department, Ben-Gurion University, Beer-Sheva 84105, Israel; guendel@bgu.ac.il}
\shortauthors{Chelouche \& Guendelman}
\shorttitle{Cosmic Stern-Gerlach Experiments}

\begin{abstract}

We show that, by studying the arrival times of radio pulses from highly-magnetized pulsars, it may be possible to detect light spin-0 bosons (such as axions and axion-like particles) with a much greater  sensitivity, over a broad particle mass range than is currently reachable by terrestrial experiments and indirect astrophysical bounds. In particular, we study the effect of splitting of photon-boson beams under intense magnetic field gradients in magnetars and show that radio pulses (at meter wavelengths) may be split and shift by a discernible phase down to a photon-boson coupling constant of $g\sim 10^{-14}\,{\rm GeV}^{-1}$; i.e., about four orders of magnitude lower than current upper limits on $g$. The effect  increases linearly with photon wavelength with split pulses having equal fluxes and similar polarizations. These properties make the identification of beam-splitting and beam deflection effects straightforward with currently available data. Better understanding of radio emission from magnetars is, however, required to confidently exclude regions in the parameter space when such effects are not observed.

\end{abstract}


\section{Introduction}

The Peccei-Quinn mechanism (Peccei \& Quinn 1977) was devised to elegantly solve to the strong-CP problem of QCD. This was accomplished by postulating a new quantum field and a new class of particles associated with it. The particles are spin-0 pseudo-scalars that couple very weakly to the electromagnetic (EM) field. It later became apparent that such particles could also provide a solution to the dark matter problem. To date, however, there is no observational evidence for the existence of such particles, dubbed axions, and it is not clear that the Peccei-Quinn solution actually works.

Besides QCD axions there are also the putative axion-like particles (ALPs). Such class of hypothetical particles has relations to dark energy and may be related to the quintessence field. These particles have been proposed as a possible solution to the apparently low opacity of the Universe to hard $\gamma$-ray radiation (e.g., de-Angelis et al. 2008) alongside other explanations (Reimer 2007, Jacob \& Piran 2008).

There is a longstanding interest in determining the physical properties of axions/ALPs. At present, laboratory experiments and astrophysical bounds imply that their  coupling constant $g<10^{-10}\,{\rm GeV}^{-1}$. Mass limits  are less stringent: if QCD axions are concerned, then their mass is probably $>10^{-6}$\,eV since otherwise the Universe would over-close, in contrast to observations. These limits, however, do not apply for ALPs. 

In recent years it has been realized that the Universe, providing us with extreme conditions, can help us constrain the axion and ALP properties. For example, the implied transparency of the universe to $\gamma$-ray radiation, if due to ALPs, puts constraints on the coupling constant and the mass. More recently, Chelouche et al. (2009; hereafter C09), have suggested a new way to detect  axions by looking at the spectra of compact astrophysical objects such as magnetars, pulsars, and quasars. Unlike terrestrial experiments where the magnetic fields are relatively small and the system size is limited by human capabilities, magnetars have magnetic fields, $B$, which are $\sim$10 orders of magnitude stronger than laboratory ones and extend over $R\sim 10$\,km scales rather than a few meters. As the photon-particle conversion probability is $\propto B^2R^2$, more stringent constraints on axion/ALP physics may be obtained.

In this paper we propose a new method to detect ALPs by utilizing the photon-ALP duality (Guendelman 2008a,b,c). As we shall show, if ALPs exist then it is possible to detect their signature in the light-curve of magnetars down to $g\sim 10^{-14}\,{\rm GeV}^{-1}$; i.e., much more sensitive than other means and complementary to the spectroscopic method of C09. This paper is organized as follows: in \S2 we lay out the formalism used to calculate first order effects in $g$ resulting in beam splitting effect that are analogous to the Stern-Gerlach experiment. Order of magnitude estimates are given demonstrating the feasibility of the method for detecting ALPs. Section 3 considers a more quantitative model pertaining to magnetars, and shows the predicted light curves when photon-ALP coupling is important.  We conclude in \S4.

\section{Splitting in in-homogenous magnetic fields}

In a recent series of papers, Guendelman (2008a,c) pointed out a photon-particle duality in a system containing  a photon-particle interaction term of the form
\begin{equation}
\displaystyle L_{\rm int} = \frac{1}{4} g \tilde{F}^{\mu \nu}F_{\mu \nu}a =  g{\bf E}\cdot {\bf B} a
\end{equation}
where $E$ is the electric field (associated with the photon), $B$ the magnetic field, and $a$ the axion field. $g$ is the coupling of particles to the EM field. The full Lagrangian for the system can be written as
\begin{equation}
L= -\frac{1}{4}F^{\mu \nu}F_{\mu \nu} -\frac{1}{2}m_\gamma^2A^2 + \frac{1}{2} \partial_\mu a \partial ^\mu a -\frac{1}{2} m_a^2 a^2 +L_{\rm int}
\end{equation}
which is the free EM  Lagrangian including an effective mass term which takes into account potential refractive index in the medium, as well as the Klein-Gordon equations for free particles ($m_a,m_\gamma$ are the particle and effective photon mass, respectively). In the absence of $L_{\rm int}$, photons and particles (e.g., axions) are well defined energy states of the system. However, once the interaction term is introduced, the only well defined energy states of the system are mixed photon-particle states. 

\begin{figure}
\plotone{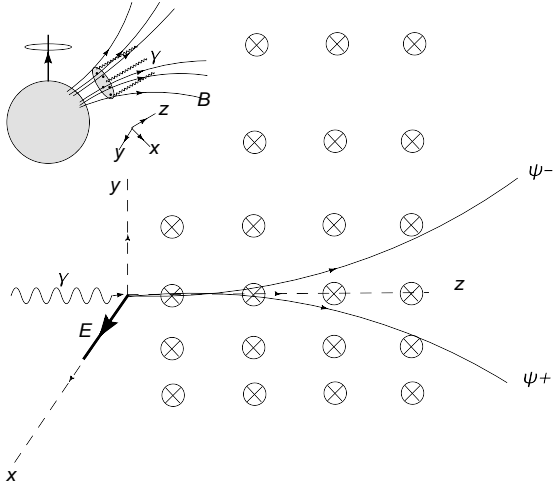}
\caption{Photon-boson beams of different "charges"  (see Appendix and Eq. \ref{psi}) would be split along magnetic field gradients in a way similar to the Stern-Gerlach experiment. Photons propagate along the $z$-axis with their polarization along magnetic field lines ($x$-axis). A Schematic view of a magnetar is also shown. Magnetic field lines originate from the magnetic pole with plasma in its vicinity  emitting beamed radiation.  One possible orientation of the coordinate system is also shown.}
\label{1}
\end{figure}

The equation of motion for the photon-particle system takes the form (e.g., C09, Raffelt \& Stodolsky 1988),\footnote{We work in natural units so that $\hbar=c=1$.}
\begin{equation}
\left [ {\bf k}^2 -\omega^2+\left \vert
\begin{array}{cc}
m_\gamma^2 & -gB_\| \omega \\
-gB_\| \omega & m_a^2
\end{array}
\right \vert \right ]  \left (
\begin{array}{l}
\gamma \\
a 
\end{array}
\right )=0,
\label{mat}
\end{equation}
where $\omega$ is the photon energy  and $B_\|$  the magnetic field in the direction of the photon polarization (the photon's $E$ field). Clearly, neither pure photon nor pure ALP states are eigenstates of the system but rather some combination of them. 

Let us now focus on the limit 
\begin{equation}
\vert m_a^2-m_\gamma(\omega)^2 \vert \ll gB_\| \omega \sim 10^{-14}g_{-14} B_{16} \lambda_m^{-1}\,{\rm eV}^2
\label{cond2}
\end{equation}
where $B_{16}=B_\|/10^{16}$\,G, $g=10^{-14}g_{-14}\,{\rm GeV}^{-1}$, and the photon wavelength, $\lambda=\lambda_m$\,m. This condition is met either near resonance where $m_\gamma^2 \simeq m_a^2$ or when both masses are individually smaller than $\sqrt{gB_\| \omega}$ (which limit is actually met is irrelevant). The eigenstates of equation \ref{mat} are then given by 
\begin{equation}
\left \vert \psi \right >_- =  \left [ \left \vert  \gamma  \right > + \left \vert a \right > \right ]/\sqrt{2},~~\left \vert \psi \right >_+ =  \left [ \left \vert  \gamma  \right > - \left \vert a \right > \right ]/\sqrt{2}
\label{psi}
\end{equation}
where $\left \vert a \right >$ is the axion state and $\left \vert \gamma \right >$ is the photon state. The eigenvalues are $m_\pm^2= \pm gB_\| \omega$. By analogy with optics, these masses are related to effective refractive indices: $n_\pm=1+\delta n_\pm\simeq 1-{m_\pm^2}/{2\omega^2}$ (for $\vert \delta n_\pm\vert  \ll1$) meaning that different paths through a refractive medium would be taken by the rays. We note that there is no dependence on the particle or photon mass so long as equation \ref{cond2} is satisfied. 

In terms of the refractive index, the equation of motion for a ray may be found by minimizing the action $\int d{\bf s} n({\bf s})$. This is completely analogous to mechanics where we substitute $\mathfrak{L} \to \omega  n_\pm$. In this case, a force is  $\partial \mathfrak{L}/\partial {\bf s} = \pm (g/2)(\partial B /\partial {\bf s})$. Using our simplified geometry, depicted in figure 1, the momentum imparted on each state is
\begin{equation}
\delta p_y^\pm = \mp  (g/2) \int dz  \left ( \partial B_x/\partial y  \right ) 
\label{dpy}
\end{equation}
where $B_x=B_x(y,z)$ (note that $dt=dz$ in the adopted units). Clearly, each of the beams will be affected in a similar way while gaining opposite momenta so that the total momentum is zero and the classical wave packet travels in a straight line (along the $z$-axis). This effect is  analogous to the  Stern-Gerlach experiment (see Fig. 1). Beam splitting effects, arising due to photon-particle (polariton) mixing, have been measured in the laboratory (Karpa \& Weitz 2006) and that an analogy exists between this case and scalar QED (see appendix).


In the limit $n_\pm \simeq 1$, the separation angle between the beams is
\begin{equation}
\theta \simeq 2p^{-1} \vert \delta p_y \vert \simeq \omega^{-1} g f_G B_\|
\label{theta}
\end{equation}
where $p$ is the beam momentum along the propagation direction, i.e., the $z$-axis. This expression holds for small splitting angles and assumes relativistic axions.  We also approximated $\int dz (\partial B_x /\partial y)= f_G B_\|$ where $f_G(N)$ is a geometrical factor depending on the magnetic field geometry, the inclination of our line-of-sight through the magnetized region (e.g., for pulsars and magnetars the magnetic field is predominantly dipolar and $f_G <1$; see \S3), and on the photon polarization.   An implicit assumption in the above expression is that the field is monotonically increasing or decreasing. In situations where the magnetic field is stochastic, $\delta p_y$ can no longer be evaluated according to equation \ref{dpy} which is linear with distance (or time) and a better treatment is that of a random walk nature whereby the average $\theta$, $\left < \theta \right > \propto \sqrt{t}$; such cases are beyond the scope of this paper and are likely to be less relevant to  compact astrophysical objects whose magnetic fields are thought to be relatively ordered. 

Assuming splitting angles $\theta \sim 10^{-2}\theta_{-2}$\,rad are detectable at radio wavelengths (see \S3), then the minimum coupling constant which can be probed, provided equation \ref{cond2} holds, is 
\begin{equation}
g_{\rm min}\sim 2\times 10^{-14} \lambda_m^{-1} f_G^{-1} B_{16}^{-1} \theta_{-2}\,{\rm GeV^{-1}}.
\label{gmin}
\end{equation}

It is important to establish that the above condition on the photon and axion mass can be materialized under  realistic conditions. Consider the non-resonant case: as the mass of ALPs is unknown, there is no reason to suspect such a condition is irrelevant. As for the photon mass, there are two contributions to the refractive index of the photon in magnetized plasma: vacuum birefringence and a plasma term (e.g., C09). The condition for the plasma term to be negligible is 
\begin{equation}
{f_G}{\theta}^{-1}  ({\omega_p}/{\omega}) ^2\ll 1.
\end{equation}
The fact that we observe a signal at photon energies $\omega$, means that $\omega \gg \omega_p$. Taking $f_G\sim 0.1$ and $\theta\sim 10^{-2}$\,rad, this condition is met if  $\omega/\omega_p > 5$ (see \S3).  We shall therefore neglect plasma effects in our analysis. A more restrictive condition is associated with the vacuum birefringence term which was calculated by Adler (1971) and used in the context of photon-particle oscillations by C09. In particular, if the refractive index $n=1+\delta n_\|^{\rm QED}$ then we demand that
\begin{equation}
2 \theta^{-1} \delta n^{\rm QED}  f_G \ll 1.
\end{equation}
For $\theta=10^{-2}$\,rad, $f_G=0.1$, and $\delta n^{\rm QED}\sim 0.05$ (C09) we get that the above ratio is of of order unity. While this condition is not strictly satisfied, we note that for slightly larger $\theta$ values (e.g., slightly lower $\omega$), this condition is met. The proper treatment of cases in which the mass terms are of order the mixing terms is beyond the scope of this paper and is the subject of a follow up investigation. In resonance, the QED and plasma contributions to the effective mass cancel; we elaborate more on these conditions in \S3.

For splitting to be observed, we require that the photon-boson beams be able to propagate through the medium hence that the refractive index is real, i.e.,
\begin{equation}
{m_\pm^2}/{\omega^2} = {gB_\|}/{\omega}\simeq {\theta}/{f_G}<1,
\label{prop}
\end{equation}
Clearly, for $\theta \gtrsim 0.1$\,rad (taking $f_G=0.1$), one beam would be attenuated as it travels through the medium. In such cases, it is more appropriate to talk about pulse shifting, as only one beam is deflected from its original path and remains observable. In this case the deflection angle, or  phase-shift, is $\tilde{\theta}=\theta/2$. 

The time-delay between the two split beams is $\delta t \simeq R\vert m_+^2-m_-^2\vert/2\omega^2 c  \ll R/c \sim 10^{-5}(R/10\,{\rm km})$\,s. Such small splitting is unlikely to be observable given the typical duration of pulses from magnetars (see below).

\section{ALPs and  the Radio Light-curves of Magnetars}

Pulsed radio emission from magnetars, whose rotation period is of order seconds, was discovered only very recently (Camilo et al. 2006) showing numerous narrow pulses over individual rotations.  The phase duration of individual pulses,  $\delta \phi \sim 10^{-2}$\,rad and so they become visible in only a small fraction of the period, $\delta_\phi\sim 10^{-3}$ (see e.g., Camilo et al. 2006 for the case of XTE\,J1810197). Given their high luminosity, the pulsed emission is likely to be beamed (with beaming factors of order $\gtrsim100$). The radio emission is known to be linearly polarized (Camilo et al. 2006).

Wishing to gain better understanding of the phenomenology of beam splitting, we note that there is no complete model for the radio emission from magnetars, and that  the magnetic field configuration is not well known. Due to the inherent uncertainties we shall consider a very simple picture of a magnetar and neglect the, possibly important, effects of magnetic loops  (Thompson \& Duncan 2001) and global field twists (Thompson et al. 2002).  In our model, the radio emission is beamed and originates from regions with high magnetic fields. The magnetic field is assumed to be dipolar and the rotation axis misaligned with  the magnetic axis (see Fig. \ref{1}).

\begin{figure}
\plotone{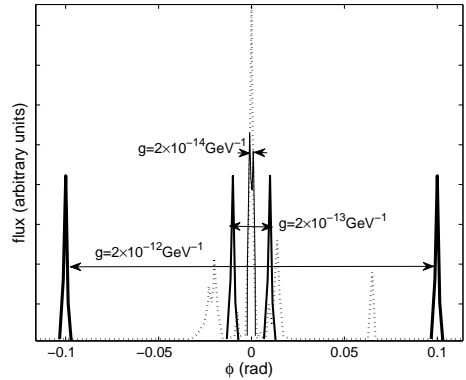}
\caption{Splitting of the peak pulse at $\phi=0$\,rad in a single rotation light-curve of XTE\,J1810197 (dotted line;  see Camilo et al. 2006) for several values of the coupling constant, $g$ and assuming $\lambda=1$\,m, $f_GB_\|=10^{15}$\,G. Note the similar fluxes of the split signals whose sum corresponds to that of the original pulse. Looking for the effects of pulse-splitting in the radio light-curves of magnetars allows one to be considerably more sensitive to light bosons compared to e.g., the CAST experiment and other astrophysical constraints. In particular, pulse splitting at meter wavelengths can be detected down to coupling constants $g_{\rm min} \gtrsim 10^{-14}\,{\rm GeV}^{-1}$ for $m_a \ll 10^{-7}$\,eV. Observing at longer wavelengths (and assuming all other parameters are fixed) will proportionally increase the sensitivity to lower values of $g_{\rm min}$ (see Eq. \ref{gmin}).}
\label{splitting}
\end{figure}

We have numerically integrated equation \ref{dpy} from the emission point of the photon near the surface of the neutron star up to large radii assuming various inclination angles of the magnetosphere with respect to the line-of-sight (photon was assumed to be linearly polarized at some direction). Performing various such integrations we choose to adopt  $f_G\sim 0.1$ and show the effect of splitting on sub-pulses in figure \ref{splitting}. Clearly, the effect is large and could be easily observed. Each pulse is split into two sub-pulses with equal fluxes (since $\left \vert \left < \gamma \vert \psi_+ \right > \right \vert^2=\left \vert \left < \gamma \vert \psi_- \right > \right \vert ^2$ when equation \ref{cond2} is fulfilled) and having a discernible phase difference.  At the CAST limit ($g\lesssim 10^{-10}\,{\rm GeV}^{-1}$; Andriamonje et al. 2007), phase shifts would be of order unity. Most notably, even for $g\gtrsim 10^{-14}\,{\rm GeV}^{-1}$, the effect can be measured at $\gtrsim$meter wavelengths. Whether or not non-split pulse components would be visible, depends on the photon polarization: splitting does not affect photons whose polarization is at right angles to the magnetic field direction. As splitting increases, asymmetric splitting, with respect to the original non-split pulse, may arise. This is caused by variations in the magnetic field gradients along different sight-lines through the magnetosphere and, in case $\vert m_\pm^2\vert /\omega^2$ is of order unity, also the somewhat different propagation speed of each beam through the medium.

The effect of splitting is wavelength dependent. In particular, $\theta \propto \lambda$ and so, by comparing the light-curve of magnetars at multiple bands, one may be able to identify split pulses from double pulses (i.e., those which are emitted as such at the source). This naturally assumes that the emission region for all bands is similar. For large enough splitting angles (and depending on $f_G$), one beam will be attenuated and only pulse shifting by a phase $\tilde{\theta}$ can be measured. To quantify the phase shift one needs to either measure it with respect to the original pulse (e.g., if some photons with a different polarization were not deflected by the effect), or statistically, by measuring typical pulse shifts between various bands and identifying the proper wavelength dependence (assuming there are no intrinsic systematic phase shifts of the pulses between the various bands).

The formalism adopted here holds equally well for the case of resonances whose importance is in the fact that they allow, in principle,  to probe more massive ALPs/axions.  As the photon mass is wavelength dependent, resonance would occur at particular frequencies, $\lambda_0$, over a limited spectral band. Once resonance bands are determined (depending on the plasma density in the magnetosphere and magnetic field strength; C09), beam splitting effects can be evaluated using equation \ref{gmin} with $\lambda=\lambda_0$. As discussed in C09, typical wavelengths for resonances which would probe more massive ALPs/axions are in the sub-mm to infrared range hence the expected splitting angles are several orders of magnitude smaller (assuming all other parameters are fixed) and the sensitivity to low values of $g$ is correspondingly lower.

\subsection{Caveats}

While we have demonstrated that the presence of finite photon-ALP coupling could lead to an easily detectable signature in the light curves of magnetars, failing to detect splitting features cannot be used to place reliable  limits on $g$ so long as the radio emission mechanism in magnetars is not well understood. For example, it might be that magnetic fields in the vicinity of the radio emitting region are considerably smaller than assumed here (e.g., if photons are emitted high above the stellar surface, in the outer parts of the magnetosphere). It could also mean that the photon polarization is less favorably inclined with respect to the magnetic field. Lastly, it may be that the radio emission is less beamed than implied by the pulse duration, in which case splitting effects would be suppressed (there are no observable implications for splitting of isotropically emitting sources).

\section{Conclusions}

\begin{figure}
\plotone{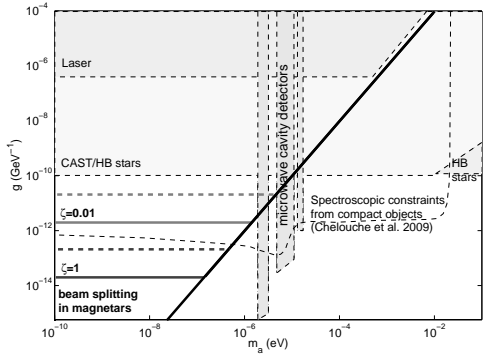}
\caption{The axion/ALP parameter space which can be explored by beam splitting in the radio light curves of magnetars as compared with other detection methods for several values of $\zeta$ [$\zeta \equiv (f_G B_\|/10^{16}\,{\rm G}) (\lambda/{\rm m})]$. Clearly, beam splitting/shifting is a potentially very efficient and sensitive method for detecting ALPs at the low mass end where equation \ref{cond2} is satisfied (resonances are not shown; see text). In particular, the sensitivities gained by this method are 3-4 orders of magnitude better than those reached by CAST and by dating of horizontal branch (HB) stars. Dashed thick lines mark the $g$-value above which one beam attenuation occurs (for each $\zeta$ and assuming $f_G=0.1$).}
\label{prm}
\end{figure}

In this paper we have described the effect of photon-ALP beam splitting in the presence of an external magnetic field -- an effect which arises solely by virtue of the interaction of the particle with the EM field. This effect is very different than (but physically related to)  that of photon-particle oscillations which were discussed in C09, and provides a complementary means for probing axion and ALP physics in magnetized environments. In this paper we focus on the observable signatures of this effect in the light curve of compact, highly magnetized objects and, in particular, the case of magnetars. The formalism developed here, and applied to magnetars, is limited to the case of negligible photon and ALP mass (relative to the interaction term) or at resonance where they equate.

We have shown that, by studying the radio light-curves of magnetars, one can be sensitive to light bosons down to very low values of the coupling constant -- about 3-4 orders of magnitude smaller than current CAST limits and other indirect astrophysical constraints. The parameter space probed by such experiments is shown in figure \ref{prm} [for various combinations of $\zeta=(f_G B/10^{16}\,{\rm G})(\lambda/{\rm m})$] and nicely complements spectroscopic searches for photon-particle oscillation features (C09). Note that, for a given $\zeta$, there is a maximum $g$ beyond which one beam is attenuated (equation \ref{prop} is not satisfied). In this case, one can probe pulse shifting, as discussed above.

The detection and verification of the splitting and shifting effects may be done by looking for a typical phase differences between pulses in the light curves of magnetars. In particular, the splitting/shifting phase would be wavelength dependent and would diminish at shorter wavelengths. In addition, the fluxes and polarizations of the split pulses would be similar. Split or shifted pulses are likely to have a different polarization than pulses whose photons do not mix with bosons.  Splitting and shifting effects could also depend on the rotation phase of the magnetar reflecting, perhaps, the configuration of the magnetic field along the line-of-sight. The various trends discussed above provide simple yet robust tests for photon-particle mixing.  It should be noted, however, that failing to detect beam splitting and shifting cannot be used to confidently exclude part of the  axion/ALP  parameter space so long as our understanding of the radio emission from magnetars is incomplete. Finally, we wish to note that the results presented here are general and apply to all (astrophysical) settings in which beamed, long wavelength emission originates from highly magnetized regions.

\acknowledgements

We are grateful to Konstantin Zioutas and the organizers of the PATRAS 4 workshop at DESY for a wonderful learning experience, and for continuous encouragement. We thank Pierre Sikivie for many discussions and extensive illuminating correspondence. We also thank Keith Baker, Giovanni Cantatore, Aaron Chou, Glennys Farrar, Yosi Gelfand, and Vicky Kaspi for valuable comments and suggestions.

\section{appendix}

Consider the scalar QED Lagrangian (neglecting mass terms and magnetic fields; Guendelman 2008a,b):
\begin{equation}
L=\partial_\mu \psi^\star \partial^\mu \psi + i e A^0\left ( \psi^\star \partial_0 \psi - \psi \partial_0 \psi^\star \right )
\end{equation}
$e$ is the charge of the scalar particle and $A^0$ the electric potential. Here $\psi=[ \left \vert a \right >+i \left \vert \gamma \right > ] /\sqrt{2}$. Using Euler-Lagrange equations we find,
\begin{equation}
\square \psi -2ieA^0\partial_0\psi =0 \to \left ( {\bf k}^2-\omega^2 -2e\omega A^0 \right ) \psi=0
\end{equation}
where the last step used an ansatz such that $\psi=e^{-i\omega t}\psi({\bf x})$. This expression is completely analogous to that given in equation \ref{mat} (after diagonalization) with $eA^0 \to gB_\|/2$. The eigenstates defined in equation \ref{psi} correspond to different charge states in the scalar QED picture and an analogue to the electric field  exerts an opposite force on each charge state leading to the beam splitting effect shown in figure 1.  

\end{document}